# Synchronous micromechanically resonant programmable photonic circuits

Mark Dong,[1,2,7] Julia M. Boyle,[1] Kevin J. Palm,[1] Matthew Zimmermann,[1] Alex Witte,[1] Andrew J. Leenheer,[3] Daniel Dominguez,[3] Gerald Gilbert,[4,8] Matt Eichenfield,[3,5,9] and Dirk Englund[2,6,10]

[1]*The MITRE Corporation, 202 Burlington Road, Bedford, Massachusetts 01730, USA*
[2]*Research Laboratory of Electronics, Massachusetts Institute of Technology, Cambridge, Massachusetts 02139, USA*
[3]*Sandia National Laboratories, P.O. Box 5800 Albuquerque, New Mexico 87185, USA*
[4]*The MITRE Corporation, 200 Forrestal Road, Princeton, New Jersey 08540, USA*
[5]*College of Optical Sciences, University of Arizona, Tucson, Arizona 85719, USA*
[6]*Brookhaven National Laboratory, 98 Rochester St, Upton, New York 11973, USA*
[7]*mdong@mitre.org*
[8]*ggilbert@mitre.org*
[9]*meichen@sandia.gov*
[10]*englund@mit.edu*

**Abstract**

Programmable photonic integrated circuits (PICs) are emerging as powerful tools for the precise manipulation of light, with applications in quantum information processing, optical range finding, and artificial intelligence. The leading architecture for programmable PICs is the mesh of Mach-Zehnder interferometers (MZIs) embedded with reconfigurable optical phase shifters. Low-power implementations of these PICs involve micromechanical structures driven capacitively or piezoelectrically but are limited in modulation bandwidth by mechanical resonances and high operating voltages. However, circuits designed to operate exclusively at these mechanical resonances would reduce the necessary driving voltage from resonantly enhanced modulation as well as maintaining high actuation speeds. Here we introduce a synchronous, micromechanically resonant design architecture for programmable PICs, which exploits micromechanical eigenmodes for modulation enhancement. This approach combines high-frequency mechanical resonances and optically broadband phase shifters to increase the modulation response on the order of the mechanical quality factor $Q_m$, thereby reducing the PIC's power consumption, voltage-loss product, and footprint. The architecture is useful for broadly applicable circuits such as optical phased arrays, 1 x $N$, and $N$ x $N$ photonic switches. We report a proof-of-principle programmable 1 x 8 switch with piezoelectric phase shifters at specifically targeted mechanical eigenfrequencies, showing a full switching cycle of all eight channels spaced by approximately 11 ns and operating at >3x average modulation enhancement across all on-chip modulators. By further leveraging micromechanical devices with high $Q_m$, which can exceed 1 million, our design architecture should enable a new class of low-voltage and high-speed programmable PICs.

## Introduction

Programmable photonics integrated circuits[1,2] (PICs) that generate periodic and spatially varying optical signals would fulfill requirements for many optical and quantum systems. For example, high-speed beam steering[3–5] in a periodic manner can provide information on the surrounding environment for range finding applications[6–9] or 3D displays[10,11]. In atom-based quantum systems, modular cluster-state architectures[12–14] rely on numerous attempts of remote-entanglement between select qubits, requiring periodic optical addressing of atoms as well as routing of spin-entangled photons. Other applications using periodic photonic switches include cycling through input vectors for optical neural networks[15] or high-throughput serial photodetection[16].

State-of-the-art PIC designs consist of meshes of reconfigurable Mach-Zehnder interferometers (MZIs), with large-scale demonstrations realized in thermo-optic[17–22], lithium niobate[23], and piezoelectric photonic platforms[24]. Several promising optical phase shifters for the construction of these mesh circuits feature microelectromechanical systems (MEMS)[25–28] with very low power consumption and scalable fabrication compatible with CMOS-foundry processes. Despite these advantages, micromechanical phase shifters[28–30] still typically require high voltages for either capacitive or piezoelectric actuation and are limited in tuning speed (~1 μs) by the presence of mechanical eigenmodes. However, if these resonances were excited purposefully at specifically designed frequencies, we would not only enhance the modulation strength of each phase shifter but enable the creation of high-speed, all-resonant programmable circuits with periodic inputs and outputs. The modulation enhancement, nominally on the order of $Q_m$, the mechanical quality factor, promises orders of magnitude decreases in each modulator's voltage-loss product, power consumption, and device footprint. The electronic control scheme (needing only sinusoids and DC biases) is also simplified, no longer requiring full arbitrary control of all phase shifts and hence the entire scope (with hardware error correction[31,32]) of SU($N$)[33,34] operations. There remains an open question whether the more focused, all-periodic actuation with mechanically resonant modulation enhancements could produce useful large-scale PICs that benefit the aforementioned applications.

Here we introduce a design architecture for optically broadband programmable PICs whose operation utilizes micromechanical resonances (Fig. 1a) for improved modulation and high-speed periodic operation in several practical configurations. We first present the theory of resonantly enhanced operation of MZI switches and direct phase modulators (Fig. 1b) for construction of phased arrays, 1 x $N$, and $N$ x $N$ photonic switches. We further apply the architecture by designing and fabricating a proof-of-principle resonantly enhanced 1x8 photonic switch. The switch consists of optical modulators based on piezoelectrically actuated cantilevers[28,35] (Fig. 1c), each engineered with target mechanical resonances for modulation enhancement (Fig. 1d). Lastly, we experimentally demonstrated successful operation of the PIC to show periodic switching of all 8 channels spaced ~11 ns apart, all while leveraging an average modulation enhancement of 3.17 (5.0 dB) compared to off-resonant, low frequency actuation. Our results demonstrate the possibility of a new class of programmable PICs operating with micromechanically resonant modulators for a broad range of applications.

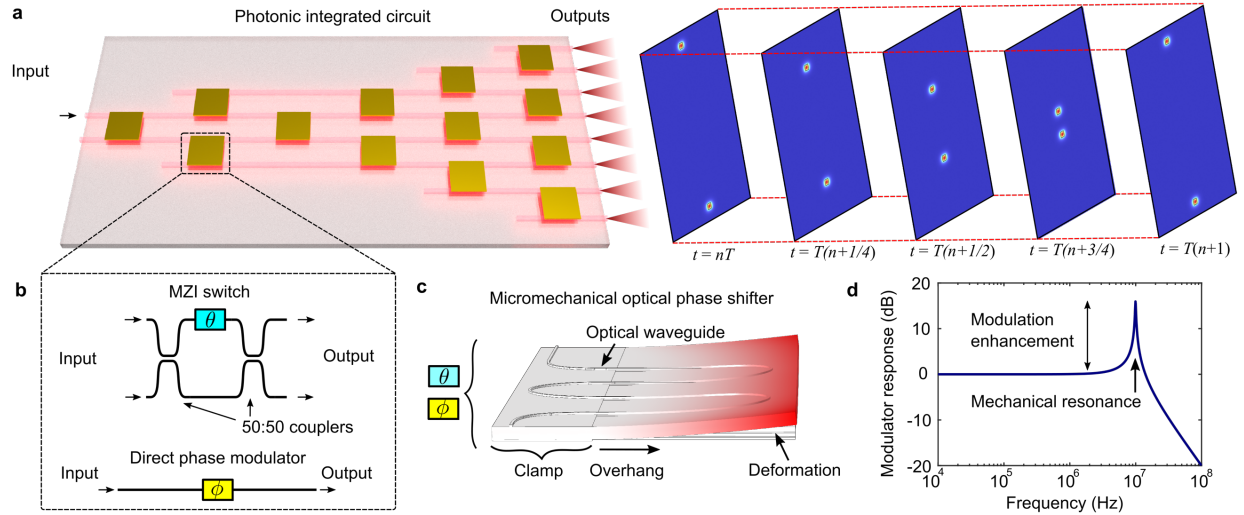

**Fig. 1: Micromechanically resonant programmable photonic integrated circuits.** a) Schematic render of a programmable photonic integrated circuit actuated with only periodic sinusoidal driving signals. The optical output channels switch periodically in time, as illustrated by the time stamps $t = nT … t = T(n+1)$. b) Diagram of circuit components, consisting of Mach-Zehnder interferometer (MZI) switches and direct phase modulators with reconfigurable phases $\theta$ and $\phi$ respectively. c) Finite-element render of a micromechanical optical phase shifter implemented as a piezo-actuated cantilever. d) Theoretical modulator response and enhancement (Equation 7) of a phase shifter with a single mechanical resonance where $\omega_0$ = 10 MHz and $Q_m$ = 40.

**Design theory and architecture**

The basic theory of a micromechanical phase modulator can be written in terms of linear transfer functions. The imparted phase as a function of the angular frequency $\omega$ is written

$$\Theta(\omega) = H(\omega)V_s(\omega) \qquad (1)$$

where $\Theta(\omega)$, $H(\omega)$, and $V_s(\omega)$ are the Fourier transforms of the imparted phase shift, modulator's response function, and modulator's applied voltage respectively. For cantilever-based modulators where the primary phase shift $\Theta$ stems from the moving boundary[28] $x$ in which $\Theta \propto x$, we derive the modulator's response function from the equivalent circuit of a piezoelectric oscillator[36] with a single resonance:

$$\Theta(\omega) \propto \frac{V_s(\omega)}{L_m} \left[ \frac{1}{i\omega R_s/L_m + i\omega R_s C_0(\omega_c^2 - \omega^2 + 2i\omega\gamma) + \omega_c^2 - \omega^2 + 2i\omega\gamma} \right] \qquad (2)$$

Here, $R_s$ is the voltage source resistance, $\omega_c = \sqrt{1/L_m C_m}$ is the modulator's series resonance frequency, $\gamma = R_m/2L_m$ is the damping coefficient, $C_0$ is the device's electrical capacitance, and $R_m$, $L_m$, $C_m$ are the motional circuit elements. For sinusoidal actuation, the applied voltage is approximated as a single frequency sinusoid of the form

$$V_s(\omega) = V_0 \pi \left[ e^{-i\phi_s} \delta(\omega - \omega_s) + e^{i\phi_s} \delta(\omega + \omega_s) \right] \qquad (3)$$

where $V_0$ is the driving amplitude, $\omega_s$ is the driving frequency, and $\phi_s$ is the phase of the driving sinusoid. Inserting Equation 3 into Equation 2 and taking the inverse Fourier transform, the time-dependent phase shift imparted by the modulator is

$$\theta(t) = \frac{1}{2\pi} \int_{-\infty}^{\infty} d\omega \, e^{-i\omega t} \Theta(\omega) \propto A_c(\omega_s) \cos(\omega_s t + \phi_s - \phi_c(\omega_s)) \qquad (4)$$

The terms $A_c(\omega_s)$, $\phi_c(\omega_s)$ are the magnitude and phase shift due to the modulator's response, respectively, defined as

$$A_c(\omega_s) = (V_0/L_m)[(\omega_c^2 - \omega_s^2 - 2\omega_s^2 \gamma \tau_0)^2 + (2\omega_s(\gamma + \gamma_s) + \omega_s \tau_0(\omega_c^2 - \omega_s^2))^2]^{-1/2} \qquad (5)$$

$$\phi_c = \tan^{-1}\left( \frac{2\omega_s(\gamma+\gamma_s) + \omega_s \tau_0(\omega_c^2 - \omega_s^2)}{\omega_c^2 - \omega_s^2 - 2\omega_s^2 \gamma \tau_0} \right) \qquad (6)$$

where the additional terms are $\gamma_s = R_s/2L_m$ and $\tau_0 = R_s C_0$. We define the modulation enhancement factor $G_m$ imparted by the micromechanical resonance as the modulation strength on or near resonance relative to the that near DC ($\omega_s \approx 0$). Assuming the source resistance is small compared to the mechanical damping and the circuit's RC response time is faster than the mechanical resonance, we calculate $G_m$ to be

$$G_m = A_c(\omega_s)/A_c(\omega_s \approx 0) = \frac{1}{\sqrt{(1-(\omega_s/\omega_c)^2)^2+(\omega_s/\omega_c)^2(1/Q_m)^2}} \qquad (7)$$

Here, $Q_m = \frac{\omega_c}{2\gamma}$ is the mechanical quality factor. When driven on resonance such that $\omega_s \approx \omega_c$, the modulation enhancement becomes equal to the mechanical quality factor: $G_m \approx Q_m$. For the design of practical large-scale PICs, these modulator variables offer important flexibility. The cantilever parameters $\omega_c$, $Q_m$ are targeted to desired values during initial design of the modulator geometry. The operating parameters $\omega_s$, $\phi_s$, and $A_s$ are then adjusted in order to maximize the modulation enhancement $G_m$ while still synchronized with the rest of the photonic circuit. See Methods for more details of the derivation.

We proceed to illustrate the design architecture of some resonantly actuated photonic circuits. A natural periodically operated device is the optical phased array, whose design and operation are shown in Fig. 2. The circuit schematic (Fig. 2a) consists of a static MZI binary tree for routing and power balancing to the $N = 8$ output channels, each with a resonantly actuated phase modulator imparting a phase $\phi_n$, defined:

$$\phi_n(t) = \pi \cos((\omega_0 + n\Delta\omega)t) \qquad (8)$$

The modulation frequency is offset by $n\Delta\omega$ depending on the channel, generating a steady phase difference between channels that periodically repeats every $T = 2\pi/\Delta\omega$. Fig. 2b shows example control voltages to the first four phase modulators, illustrating the phase differences plotted over time. The total modulation enhancement will be strongest when the mechanical eigenfrequency $\omega_{cn}$ match the modulation frequency of the $n$th channel

$$\omega_{cn} = \omega_0 + n\Delta\omega \qquad (9)$$

Once the phases are set, the output optical waveguides simply need to be routed, spaced, and coupled off-chip to complete the beam scanner[37]. We numerically simulated the 1D phased array whose optical outputs are modeled as edge-emitting waveguides to show device operation. Using the phases defined in Equation 8, the simulation calculates the Gaussian beam equation modulated in time

$$E_n(x, z, t) = E_0(1/q(z)) \exp\left(-ik(x - x_n)^2/(2q(z))\right) \exp\left(ikz + i\phi_n(t)\right) \qquad (10)$$

where $E_0$ is the field amplitude, $x_n$ is the $n$th output waveguide's origin point, $q(z)$ is the complex beam parameter given by $\frac{1}{q(z)} = \frac{1}{R(z)} - i\frac{\lambda}{\pi w(z)^2}$, $R(z)$ is the beam curvature whose inverse is given by $\frac{1}{R(z)} = \frac{z}{z^2 + z_R^2}$, $w(z) = w_0\sqrt{1 + z^2/z_R^2}$ is the propagation-dependent beam waist, $w_0$ is the initial beam waist at $z = 0$, $z_R = \pi w_0^2/\lambda$ is the Rayleigh range, and $\lambda$ is the wavelength. The spacing between each $x_n$ is a half wavelength and the initial beam waist is set to $w_0 = 0.5$. The total intensity is then given by

$$I(x, z, t) = |\sum_{n=1}^{8} E_n(x, z, t)|^2 \qquad (11)$$

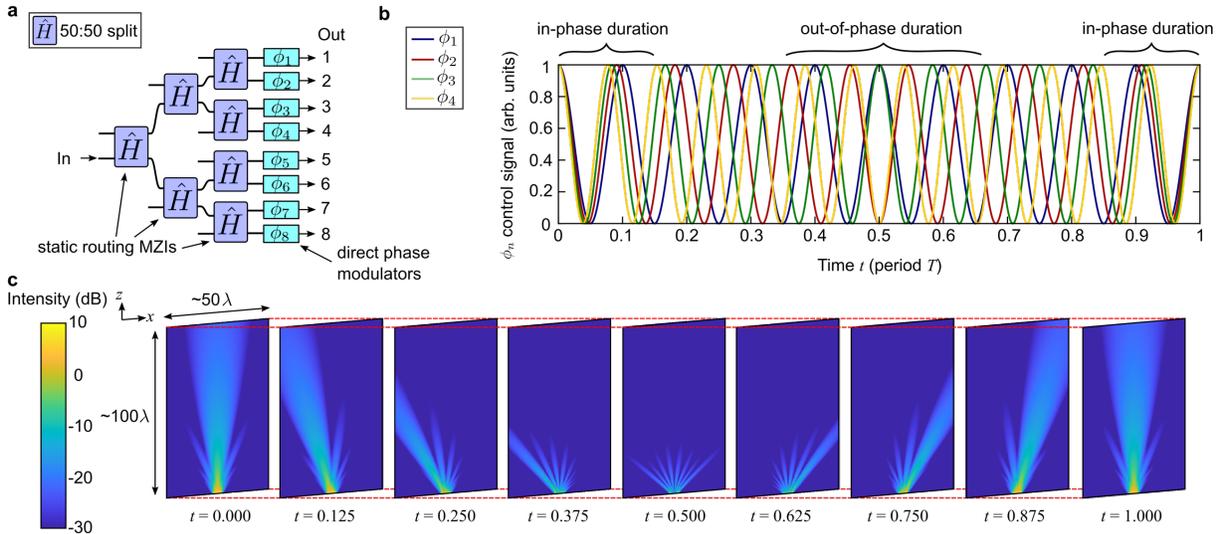

**Fig. 2: Resonantly actuated 8-channel phased array.** a) Architecture of a PIC-based phased array based on 8 optical outputs consisting of static MZIs for power routing and direct phase modulators at the output of each channel. These modulators offset each channel's phase by a steady rate over time, defined by the frequency difference $N\Delta\omega$. b) Example plots of the control signals applied to the first four channels $\phi_1 - \phi_4$. The channels periodically drift in and out of phase on a timescale defined by $T = 2\pi/\Delta\omega$. c) Gaussian beam calculations of the 8 output phased array at discrete times over a spatial domain of $50\lambda$ x $100\lambda$ of the $x$ and $z$ axes respectively. The central beam combines constructively when all control phases $\phi_n$ are in phase ($t = 0$). At subsequent times, the beam sweeps counterclockwise and repeats at every $t = T$.

Fic. 2c plots the output intensity (Equation 11) at different times $t$, showing a periodic beam-scanner actuated with the sinusoidal driving signals corresponding to Fig. 2b. The in-phase and out-of-phase durations illustrate how the outputs interfere constructively or destructively in order to steer the beam.

Another class of photonic circuits that benefits from periodic actuation is the integrated photonic switch. The basic 1 x $N$ switch may be implemented as a mesh of Mach-Zehnder interferometers (MZIs), as shown in Fig. 3a. Input light $E_{in}$ is coupled through a single port on the left and routed to the $n$th output $E_n$ on the right with the transfer matrix $T_{ij}(\theta_{ij})$ applied to each depth of MZIs. The normalized output intensities $I_n$ can be written as a product of sinusoids

$$I_n = \prod_{m=1}^{M}(1 + s(m,n)\cos(2\theta_{ij}^{(m,n)})) \qquad (12)$$

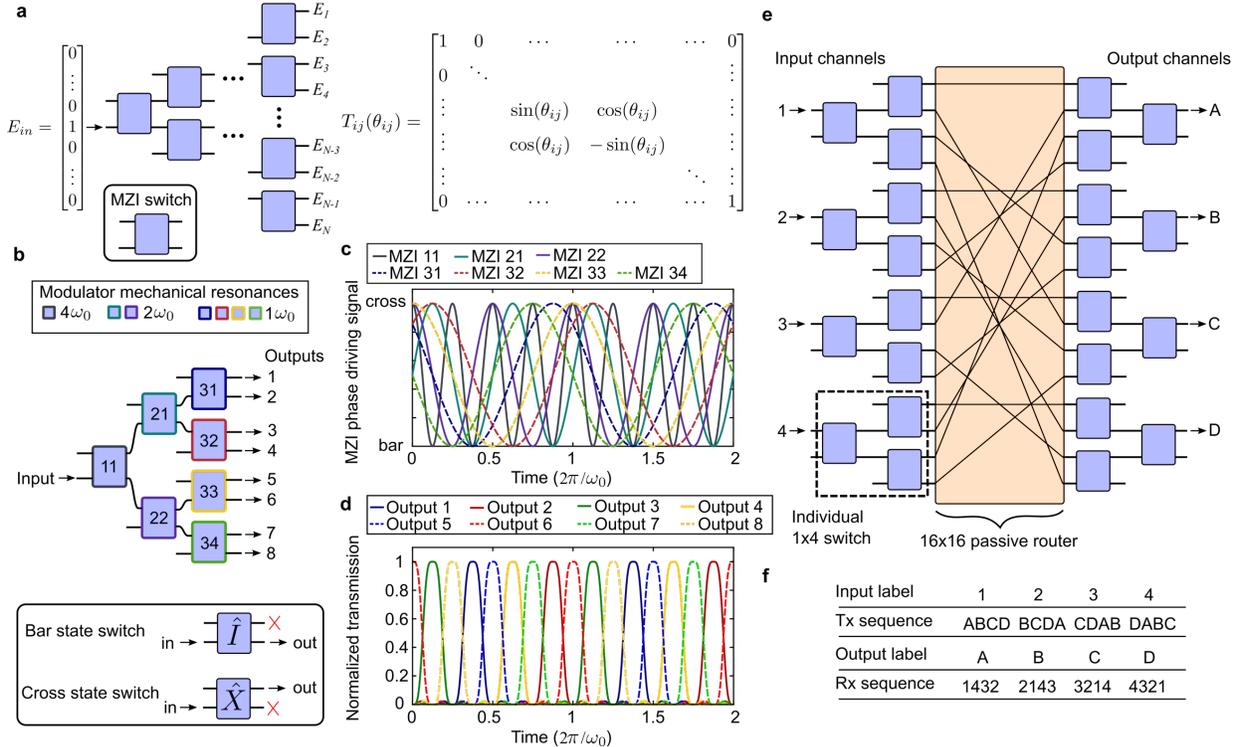

**Fig. 3: Architecture of micromechanically resonant 1 x $N$ and $N$ x $N$ integrated optical switches.** a) Schematic of a general 1 x $N$ optical switch implemented as a binary tree mesh of MZIs with transfer function $T_{ij}$. b) Design of a resonantly actuated 1 x 8 optical switch operating at three distinct frequencies, each unique to a column of MZIs. c) Examples of driving signals in time domain to each of the MZIs with appropriate phase offsets. Each signal oscillates between bar and cross switching states. d) Theoretical normalized optical intensity of the 8 output channels plotted in time domain when the circuit in b) is driven with the signals in c). e) Schematic of a resonantly actuated 4 x 4 matrix switch consisting of 8 individual 1 x 4 switches connected through a 16 x 16 passive router. The individual switches are programmed with different driving signals to facilitate a proper switching order. f) Example periodic transmission and receiving pattern through the 4 x 4 matrix switch.

Here, $M = \log_2(N)$ the total circuit depth, $s(m,n) = \pm 1$ a sign that depends on whether the $n$th output channel's optical path is connected to the $m$th depth MZI's bar or cross port, and $\theta_{ij}^{(m,n)}$ is the time-dependent phase setting of that MZI. We illustrate a specific example of operating a resonantly actuated 1 x 8 switch (Fig. 3b). In this circuit, there are 7 MZIs labeled with the depth and the row, each with a modulator having a mechanical eigenfrequency that is an integer multiple of a base frequency $\omega_0$. All MZIs are actuated synchronously with a resonant sine wave and phase offset (Fig. 3c) such that all light is periodically routed to a single channel at specific moments in time. Fig. 3d plots Equation 12 with $N = 8$ and $M = 3$ driven by the time-dependent $\theta_{ij}^{(m,n)}$ signals in Fig. 3c, showing a periodic train of pulses emitted by each channel in sequence. The design of the optical switching alternates between the top half and the bottom half of the tree whose order can be adjusted by changing the driving phase offsets.

In addition to 1 x $N$ switches, resonantly actuated $N$ x $N$ matrix switches are also possible. One representative construction of $N = 4$ is shown in Fig. 3e and generally consists of $2N$ number of 1 x $N$ switches and a connecting $N$ x $N$ static optical router. The individual 1 x $N$ switches are programmed such that in one actuation period, each input channel appears at every output channel of the $N$ x $N$ matrix switch. In other words, by adjusting the static connections and the sinusoid phases of the 1 x $N$ switch, the matrix switch can be tailored to form a transmission (Tx) and receiving (Rx) sequence to suit the desired application. Fig. 3f displays a table of one possible Tx and Rx sequence. The Tx sequence entries are how each input channel transmits to the corresponding output in the order of the output channels labeled. Conversely, the Rx sequence entries are the order of the input channels received by the labeled output channel. In $N = 4$ cycles, all inputs are transmitted once to all outputs.

**Micromechanically resonant 1 x 8 photonic switch**

To demonstrate micromechanically resonant architecture, we designed, fabricated, and characterized a proof-of-principle 1 x 8 photonic switch. The device is realized on a PIC platform whose layer stack consists of integrated silicon nitride waveguides and piezoelectric aluminum nitride, all fabricated on 200 mm silicon wafers[24,38]. Fig. 4a illustrates the switch's resonant MZI design based on variable overhang cantilevers[28]. Each MZI consists of two independently controllable phase shifters: the first is used for DC biasing to the mid-point of the MZI response function; the second is driven sinusoidally between the approximate bar and cross states while taking advantage of any modulation enhancement from the mechanical eigenmode. We targeted cantilevers with mechanical eigenmodes at approximately 10 MHz, 20 MHz, and 40 MHz (corresponding to a base frequency of $\omega_0 \approx 2\pi \times 10$ MHz), all arranged according to the schematic in Fig. 3b. After fabrication, we packaged the photonic switch on a custom PCB with high-speed voltage amplifiers driven with multi-channel arbitrary waveform generators (AWGs) for electrical inputs. Optical inputs were fed through a fiber array while optical outputs were collected via a free-space imaging system for characterization. Fig. 4b shows an optical microscope image of the fully processed photonic switch with labeled MZIs.

To program the circuit, we calibrated all MZIs as follows. We first measured the responses of all the DC phase shifters to store the target midpoint DC voltages. Next, we measured the small-signal, linearized response of all AC phase shifters around the DC bias point using a vector network analyzer. Fig. 4c-e plots the AC actuation response of each column of MZIs averaged 100 times and normalized to the response at low frequency (100 kHz), showing clear modulation enhancement (peaks at roughly 5 dB - 10 dB) near our desired frequencies of interest. Despite the eigenfrequencies not precisely matching up for all MZIs due to the presence of other mechanical modes, variability in device release, and local stress gradients, the wide range of enhancement allows flexibility in choosing the base frequency $\omega_0$. Once the base frequency was chosen, we calibrated the AC $V_\pi$, the voltage required to drive the MZI between its bar and cross states with a particular modulation enhancement. Lastly, we set and synchronized the phases $\phi_s$ of all driving signals for periodic switching operation. The PIC technology's high degree of long-term stability[24] allowed our calibration settings to persist and the photonic switch to operate in open loop over several weeks.

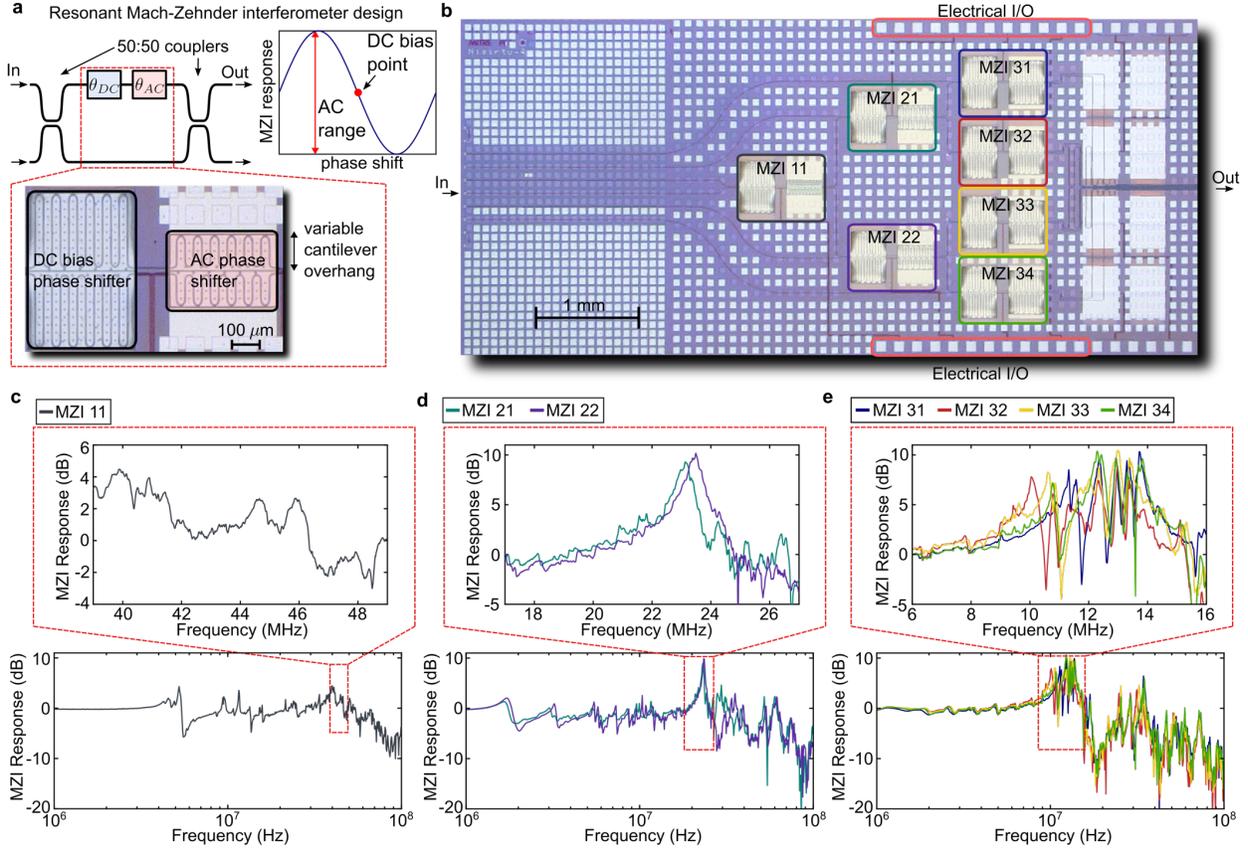

**Fig. 4: Design and characterization of a piezoelectric 1 x 8 integrated optical switch.** a) Illustration of the resonantly actuated MZI design with two independent phase shifters: one larger, non-resonant cantilever for static DC biasing to the midpoint of the MZI's sinusoidal response function and one smaller, AC cantilever designed with a target mechanical eigenfrequency. b) Optical microscope image of the fully fabricated PIC with labeled MZIs, optical inputs and outputs, and electrical bonding pads. c)-e) Measured small-signal response of the AC cantilever for the first (MZI 11), second (MZI 21 and 22) and third (MZI 31, 32, 33, and 34) columns respectively - each measured trace is averaged 100x. In this device, the targeted base frequency is nominally 10 MHz; zoomed-in plots show the response near 40 MHz, 20 MHz, and 10 MHz respectively.

We measured the performance of the fully programmed photonic switch for a particular base frequency $\omega_0 = 11.3$ MHz, as shown in Fig. 5. This specific base frequency was selected according to the isolation of the driven mechanical eigenmode as well as the overall modulation enhancement (5.0 dB on average from Fig. 4c-e) of all 7 MZIs. Fig. 5a plots time-resolved measurements using a high-speed photodiode, normalized for different losses and collection efficiencies, of all 8 output channels. The traces show clear peaks every ~11 ns which follow an 8-channel switching order that cycles every period $T = 2\pi/\omega_0$. We also plot the AC driving voltages of all MZIs (Fig. 5b) and list the values of their amplitudes $V_0$ and phase $\phi_s$ (Fig. 5c). We note that the relative phases of the driving signals differ from the ideal values in Fig. 3c as not all MZIs are driven precisely at the resonance peak – there are adjustments to $\phi_s$ to compensate for the induced, frequency-dependent phase $\phi_c$ (Equation 6) of every individual cantilever. The switching fidelity and ordering of all channels generally matches the theoretical case as shown in Fig. 3d. However, while the maximum switching contrast during times the channels are "off" is approximately 0.032 +/- 0.007 (14.95 dB) for all channels, some channels show non-negligible pulsing during off-peak times (output 2 is a notable example in Fig. 5a). These off-peaks occur due to imperfect calibration on the larger-overhang cantilevers, which occasionally exhibit mechanical nonlinear coupling (see Supplementary Methods) over many periods between different eigenmodes.

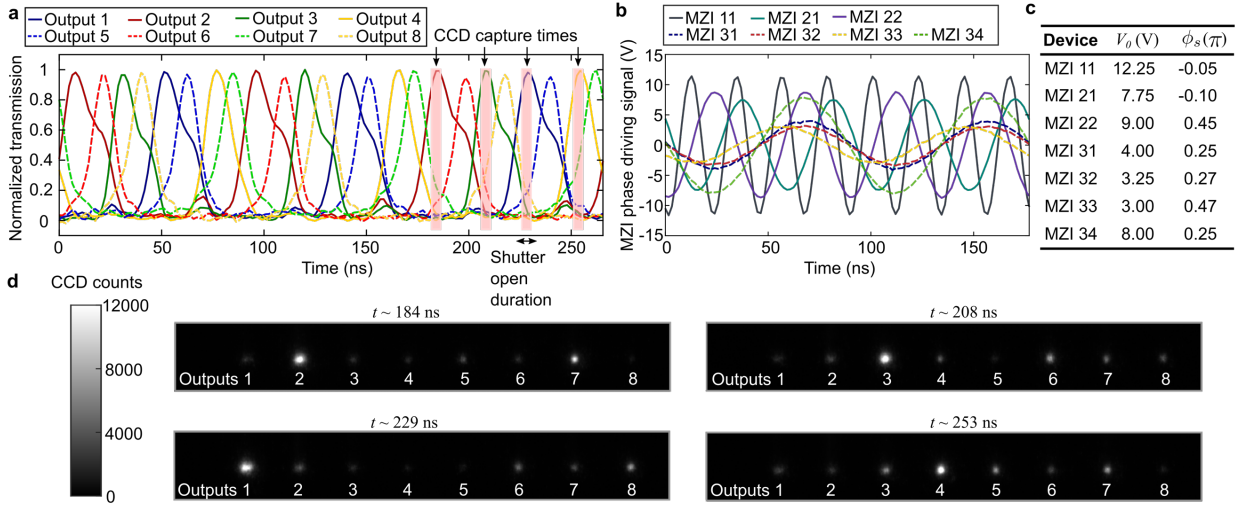

**Fig. 5: Measurements of the 8-channel full switching sequence**. a) Time-resolved measurements of all 8 optical outputs during a full switching sequence with a base frequency of 11.3 MHz for three full periods. CCD capture times for the imaging experiment in d) are also labeled, depicting the small integration region defined by the camera shutter. b) Plots of the sinusoidal driving inputs to the all seven MZIs. c) Table of driving parameters, including voltage amplitude $V_0$ and phase $\phi_s$. The driving phase term compensates for the variation in each cantilever's phase response d) CCD images captured at various times $t$ as labeled in part a). Each image integrates the output from all 8 optical channels for 4.8 ns and illustrates an unnormalized snapshot of the output at different temporal offsets in the total switching sequence.

We further characterized the switching sequence by imaging all 8 waveguide outputs simultaneously onto an intensified charge-coupled device (ICCD). The camera's image intensifier, acting as a high-speed shutter, was programmed synchronously with the MZI's driving sinusoids, allowing the capture of a small temporal window of light as illustrated in Fig. 5a. These capture times were precisely aligned to the output switching sequence by an analog gate signal generated from the same multi-channel AWGs driving the photonic switch. Fig. 5d shows four CCD images at different times $t$ (relative to Fig. 3a) of a 4.8 ns exposure of the 8 output channels. Each image shows the charge count during a small subperiod (4.8 ns) of the full switching sequence, accumulated over three periods and amplified to improve the signal-to-noise ratio. The images are a measure of the unnormalized intensities of the switch's output, with bright channels matching those in the photodiode trace. Both the imaging and time-resolved experiments demonstrate the operation and viability of our 1 x 8 resonantly actuated photonic switch. See Supplementary Methods for more details of the device calibration and imaging experiment.

## Discussion

We have shown theoretically and experimentally the concept of mechanically resonant modulation for the construction of periodically operated programmable photonics. While the architecture encompasses a broad range of PIC designs and applications, there are several future improvements to the current experimental realization. The cantilever modulators' occasional nonlinear mechanical coupling is likely due to uneven undercuts and film stresses during fabrication. This can be mitigated by shortening the cantilever overhang to facilitate better undercutting and patterning the electrodes to more efficiently excite the eigenmodes of interest. Furthermore, the current photonic switch only relied on the fabricated geometry for targeting the value of the mechanical eigenmode. Thus, the overall achieved modulation enhancement fell short of the measured maxima due to small mismatches in mechanical eigenfrequencies. The tuning of mechanical eigenfrequencies across all devices on the PIC may be improved by post-process trimming and ablation[39] to adjust the mechanical spring constant or mass to optimize all cantilevers to the previously demonstrated[28] $Q_m = 40$. Lastly, the overlap of output channel intensities in the full switching sequence (Fig. 5a) is improved by adding additional frequencies to sharpen the output pulses (see Supplementary Discussion).

More broadly, the infusion of ideas from micro- and nano-mechanical engineering of high-Q oscillators[40] as well as high-Q electronic oscillators should further decrease the operating voltage of these photonic circuits by many orders of magnitude. Structures have been demonstrated in thin-film piezoelectric materials with MHz eigenfrequencies that have mechanical quality factors exceeding 10,000[41,42]. Other MEMS platforms[43] have achieved even higher quality factors of >$10^6$ with well-designed structures. Due to the simplification of the PIC's electrical control, commercially available electronic tank circuits can now be made to match each mechanical resonance, adding another multiplicative reduction in voltage. By leveraging MEMS with integrated photonics, a new category of programmable PICs with ultra-low power consumption, footprint, and losses should be possible.

**Acknowledgements**

Major funding for this work is provided by MITRE for the Quantum Moonshot Program. D.E. acknowledges partial support from Brookhaven National Laboratory, which is supported by the U.S. Department of Energy, Office of Basic Energy Sciences, under Contract No. DE-SC0012704 and the NSF RAISE TAQS program. M.E. performed this work, in part, with funding from the Center for Integrated Nanotechnologies, an Office of Science User Facility operated for the U.S. Department of Energy Office of Science. M.D. and M.Z. thank L. Chan, K. Dauphinais, and S. Vergados for their support in constructing and testing the mechanical and electronic components.

**Author contributions**

M.D., J.M.B., K.J.P., M.Z., and A.W. built the experimental setups and performed the experiments. M.D. performed the theoretical analysis and designed the photonic integrated circuit. M.E. and A.J.L., with assistance from D.D., supervised the device fabrication. M.Z. and M.D. designed and programmed the electronic control system. M.D. and D.E. conceived the programmable PIC architecture. G.G., M.E., and D.E. supervised the project. M.D. wrote the manuscript with input from all authors.

**Additional information**

Supplementary information is available for experimental methods related to programming and calibrating the photonic integrated circuit.

**Competing interests**

D.E. is a scientific advisor to and holds shares in QuEra Computing. The authors declare no other competing interests.

**Data availability**

The data that support the plots within this paper are available from the corresponding authors upon reasonable request.

## Methods

**Linear response function of a piezoelectric optical modulator.** We model the response function of our piezoelectrically actuated optical modulators using the standard equivalent circuit of a piezoelectric device[36] (Butterworth-van-Dyke model). The circuit has a frequency-dependent effective impedance $Z_{eq}(\omega)$ given by

$$Z_{eq}(\omega) = R_s + \frac{\omega_c^2 - \omega^2 + 2j\omega\gamma}{j\omega/L_m + j\omega C_0(\omega_c^2 - \omega^2 + 2j\omega\gamma)} \quad (M1)$$

where $j$ is the imaginary unit, $R_s$ is the series source resistance, $\omega_c = \sqrt{1/L_m C_m}$ is the series resonance frequency, $\gamma = R_m/2L_m$ is the damping coefficient, $C_0$ is the device's electrical capacitance, and $R_m, L_m, C_m$ are the motional circuit elements. The frequency-dependent applied voltage $V_s(\omega)$ and current $i(\omega)$ are defined

$$V_s(\omega) = V_0 \pi \left[ e^{-i\phi_s} \delta(\omega - \omega_s) + e^{i\phi_s} \delta(\omega + \omega_s) \right] \quad (M2)$$

$$i(\omega) = V_s(\omega)/Z_{eq}(\omega) \quad (M3)$$

We are interested in the charge accumulated on the motional capacitor $q_m$, which is analogous to the motional displacement $x$ responsible for the optical phase shift. We make the assumption that the phase shift $\theta$ ultimately imparted by the modulator linearly follows the displacement, i.e. $x = B_1 q_m$ and $\theta = B_2 q_m$ where $B_1, B_2$ are constants representing the optomechanical coupling. We calculate the current flowing through the motional circuit elements:

$$i_m(\omega) = i(\omega) \frac{1/j\omega C_0}{1/j\omega C_0 + j\omega L_m + R_m + 1/j\omega C_m} \quad (M4)$$

and using the relation $i_m(\omega) = j\omega q_m(\omega)$, we find

$$q_m(\omega) = \frac{V_s(\omega)}{L_m} \left[ \frac{1}{j\omega R_s/L_m + j\omega R_s C_0(\omega_c^2 - \omega^2 + 2j\omega\gamma) + \omega_c^2 - \omega^2 + 2j\omega\gamma} \right] \quad (M5)$$

To find the time-dependent motional displacement, we first introduce $\gamma_s = R_s/2L_m$ and $\tau_0 = R_s C_0$ for simplicity. Inserting Equation M2 into M5 and performing the inverse fourier transform, we calculate the time-dependent charge to be

$$q_m(t) = \frac{1}{2\pi} \int_{-\infty}^{\infty} d\omega e^{-j\omega t} q_m(\omega) = A(\omega_s) \cos(\omega_s t + \phi_s + \phi_c(\omega_s)) \quad (M6)$$

$$A_c(\omega_s) = (V_0/L_m)[(\omega_c^2 - \omega_s^2 - 2\omega_s^2 \gamma \tau_0)^2 + (2\omega_s(\gamma + \gamma_s) + \omega_s \tau_0 (\omega_c^2 - \omega_s^2))^2]^{-1/2} \quad (M7)$$

$$\phi_c = \tan^{-1}\left( \frac{2\omega_s(\gamma + \gamma_s) + \omega_s \tau_0 (\omega_c^2 - \omega_s^2)}{\omega_c^2 - \omega_s^2 - 2\omega_s^2 \gamma \tau_0} \right) \quad (M8)$$

We define the modulation enhancement factor $G_m$ by normalizing the actuation amplitude $A_c(\omega_s)$ to that near DC ($A_c(\omega_s \approx 0)$).

$$A_c(\omega_s \approx 0) = \frac{V_0}{\omega_c^2 L_m} \quad (M9)$$

$$G_m = A_c(\omega_s)/A_c(\omega_s \approx 0)$$
$$= \left[ \left(1 - \left(\frac{\omega_s}{\omega_c}\right)^2 - \left(\frac{\omega_s}{\omega_c}\right)^2 \frac{\omega_c \tau_0}{Q_m}\right)^2 + \left(\frac{\omega_s}{\omega_c Q_m}(1 + R_s/R_c) + \omega_s \tau_0 (1 - \left(\frac{\omega_s}{\omega_c}\right)^2)\right)^2 \right]^{-1/2} \quad (M10)$$

where we have inserted $Q_\mathrm{m} = \omega_\mathrm{c}/2\gamma$ the mechanical quality factor. If we assume operating conditions of i) small source resistance compared to the mechanical damping ($R_\mathrm{s} \ll R_\mathrm{m}$) and ii) the mechanical resonance is far below the circuit's RC response ($\tau_0\omega_\mathrm{c}, \tau_0\omega_\mathrm{s} \ll 1$), then the modulation enhancement simplifies considerably to

$$G_\mathrm{m} = \frac{1}{\sqrt{(1-(\omega_\mathrm{s}/\omega_\mathrm{c})^2)^2 + (\omega_\mathrm{s}/\omega_\mathrm{c})^2(1/Q_\mathrm{m})^2}} \qquad (\mathrm{M11})$$

which is Equation 7 in the main text.

# Synchronous micromechanically resonant programmable photonic circuits: Supplementary Information

Mark Dong,[1,2,7] Julia M. Boyle,[1] Kevin J. Palm,[1] Matthew Zimmermann,[1] Alex Witte,[1] Andrew J. Leenheer,[3] Daniel Dominguez,[3] Gerald Gilbert,[4,8] Matt Eichenfield,[3,5,9] and Dirk Englund[2,6,10]

[1]*The MITRE Corporation, 202 Burlington Road, Bedford, Massachusetts 01730, USA*
[2]*Research Laboratory of Electronics, Massachusetts Institute of Technology, Cambridge, Massachusetts 02139, USA*
[3]*Sandia National Laboratories, P.O. Box 5800 Albuquerque, New Mexico, 87185, USA*
[4]*The MITRE Corporation, 200 Forrestal Road, Princeton, New Jersey 08540, USA*
[5]*College of Optical Sciences, University of Arizona, Tucson, Arizona 85719, USA*
[6]*Brookhaven National Laboratory, 98 Rochester St, Upton, New York 11973, USA*
[7]mdong@mitre.org
[8]ggilbert@mitre.org
[9]meichen@sandia.gov
[10]englund@mit.edu

## 1. Supplementary Methods

**Experimental setup for characterizing the PIC.** Supplementary Fig. 1 depicts the experimental setup for the resonant tree calibration and qualification. We used a Ti:Sapphire laser at 737 nm wavelength for the 1 x 8 photonic switch characterization. The light was coupled to the PIC from a fiber aligned to an on-chip grating. The input polarization was controlled with a half and quarter waveplate in order to maximize transmission into the TE mode of the SiN waveguide. The eight edge-coupled outputs of the switch were collected with a high-NA objective (0.9 NA) and refocused onto either an intensified charge coupled device (ICCD) or a 125 MHz photodiode. The collected light was filtered with a free-space linear polarizer before detection.

For electrical control, the photonic switch was cleaved and wire bonded to a custom printed circuit board (PCB). The cantilever phase shifters were electrically driven with two 625 MS/s arbitrary waveform generator (AWG) PXIe cards, each with four output channels with a voltage range of ± 2.5 V and 5x amplified on the PCB at a max slew rate of 8000 V·μs$^{-1}$. The cantilevers were operated in push pull configuration in order to maximize actuation. A laptop computer controlled all of the electrical equipment and sent programmed pulse sequences to the voltage controller, which then synchronously actuated all devices on the photonic switch. The computer then collected all data from the oscilloscope and camera for post-processing.

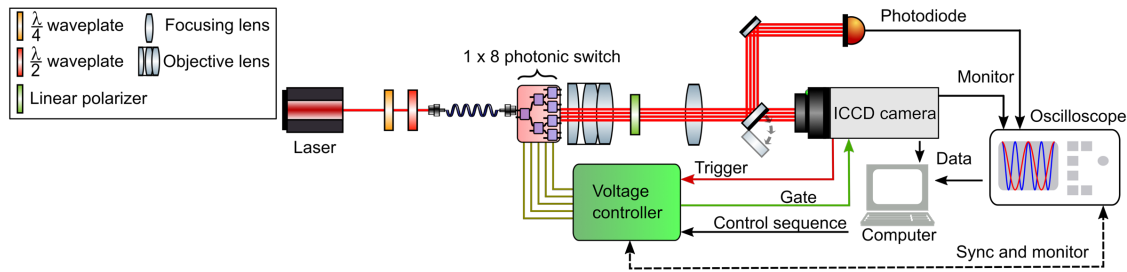

**Supplementary Fig. 1: Experimental setup for characterization of the 1 x 8 photonic switch.** Diagram of the components used in the experiment characterizing the photonic switch.

**Timing and triggering of voltage controller and ICCD.** Supplementary Fig. 1 illustrates several timing signals between the computer, voltage controller, and ICCD camera. These signals from an exemplary exposure of the ICCD are plotted in Supplementary Fig. 2. In our high-speed imaging experiments, we first programmed all photonic control signals into the memory buffer of the voltage controllers. We then set the ICCD to start the exposure but with the shutter off initially. The "fire" monitor on the camera, which goes high when exposure is active (Supplementary Fig. 2a), was used to trigger the voltage controllers to execute the synchronized sinusoids to the PIC. In addition to controlling the PIC, the voltage controller also sent a synchronous "gate" signal to directly turn on the camera shutter at specific moments in time relative to the firing of the photonic switch. Only when the "gate" and "fire" signals are both on (TTL high) will the camera measure the optical outputs. We subsequently adjusted the timing offset of the gate signal to generate different images of the 8-channel output (Supplementary Fig. 2b, c). We see three distinct gate signals (4.8 ns width) per image capture to improve the signal-to-noise ratio, spaced at 50 times the fundamental frequency period (~4.425 μs). We note the spacing is limited by the ICCD direct-gate repetition limit of 200 kHz.

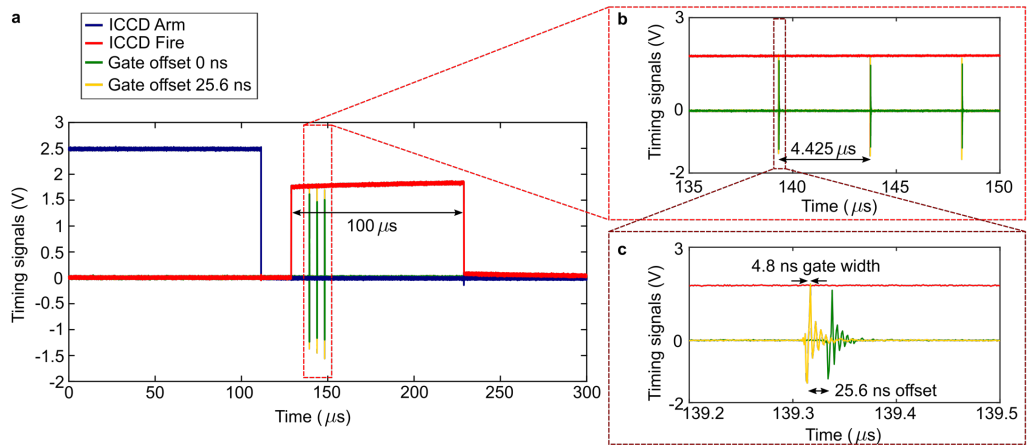

**Supplementary Fig. 2: Timing signals for the ICCD.** a) Monitored voltages from the ICCD's "arm" and "fire" channels, indicating when the camera starts the exposure. Two "gate" monitor signals from two separate image captures, indicating when the shutter has opened, are also plotted for comparison. b) Zoomed-in portion of the gate signals, showing three shutter opens per exposure. c) Further zoomed-in plot of the two gate monitor signals showing the widths and relative programmed delays between image captures.

**Device calibration and mechanical stability.** We calibrated the photonic switch in sequence, storing all voltage values in a file that could be reloaded for each experiment. We first calibrated the DC voltages, specifically the 50:50 operating point, starting with MZIs in column 3 and proceeding to columns 2 and then 1. Next, we performed a small-signal frequency response measurement of the AC modulators (Fig. 4c-e in the main text) using a two-port vector network analyzer. We connected port 1 to the desired target device and port 2 to the photodiode output. By measuring the scattering parameters, specifically S21, we obtained the MZI response. From these plots, we then selected the base frequency (11.3 MHz in the main text) to optimize the overall modulation enhancement of all devices. At 11.3 MHz, the modulation enhancement factors were 1.4, 3.9, 2.9, 6.9, 3.2, 1.7, and 2.2 for MZIs 11, 12, 22, 31, 32, 33, 34 respectively, averaging an overall 3.17x or 5.0 dB for the whole circuit. Lastly, we proceeded to optimize the output pulse sequence by adjusting the AC sinusoid amplitude and phase. Supplementary Fig. 3 shows the phase synchronization process, where we monitored the full switching sequence and adjusted the phases to achieve the desired switching order and compensate for the phase $\phi_c$ of each cantilever.

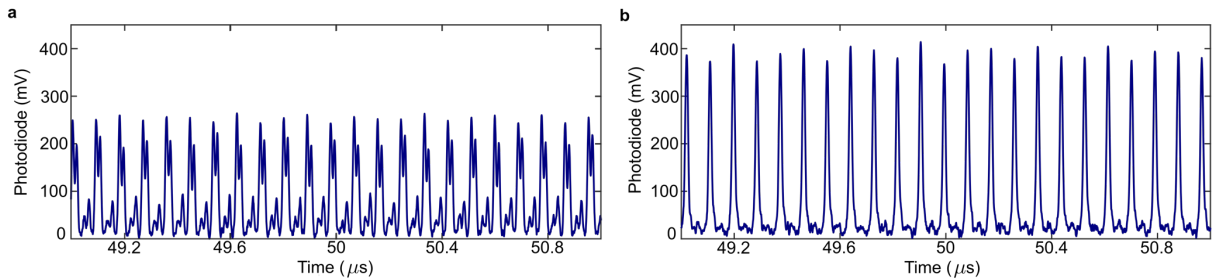

**Supplementary Fig. 3: Phase alignment and compensation of MZIs.** a) Time-resolved output of optical channel 6 before the MZI has undergone the calibration procedure. b) Time-resolved output of optical channel 6 after appropriate amplitude, offset, and phase values are implemented on the MZI.

During the calibration procedure, we note that not all AC cantilevers were able to perfectly hit the on-off switching range. Specifically, the longer overhang cantilevers (all in column 3 MZIs) exhibited some mechanical instability when driven at high voltages. Supplementary Fig. 4 plots this phenomenon for two MZIs: a more stable MZI 34 and a less stable MZI 31. We applied digital signal processing techniques on the optical modulation resulting from each MZI and plotted the time-dependent amplitudes of all relevant harmonics. The dynamics of the less stable modulator (Supplementary Fig. 4d) suggests nonlinear mechanical coupling. To alleviate this effect, we applied lower voltage sinusoids (Fig. 5c in the main text) for better stability at the cost of lower switching fidelity. This nonlinear coupling does not occur on the shorter overhang cantilevers (MZIs in columns 1 and 2). We plan to study in more detail these mechanical effects in future work.

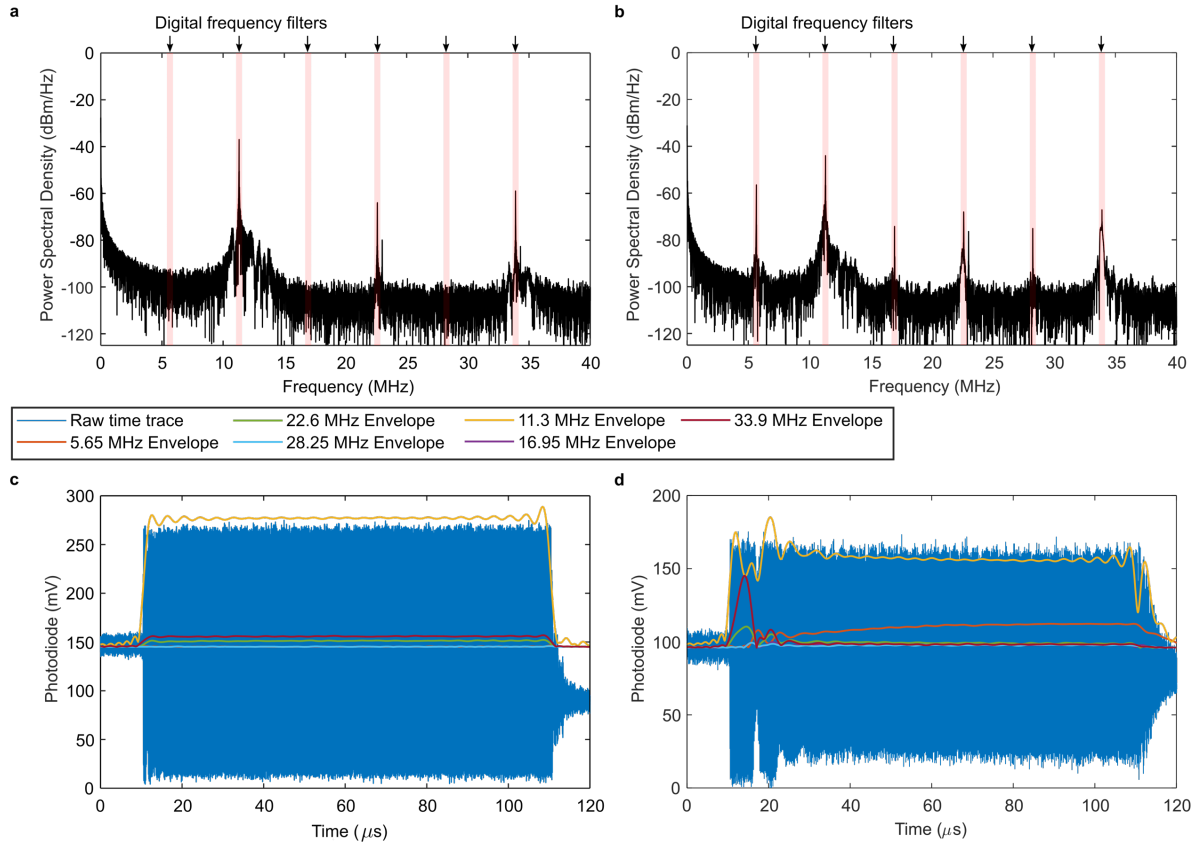

**Supplementary Fig. 4: Mechanical harmonic generation and frequency drift in MZIs.** Frequency and time-domain plots of the optical modulation when MZIs 31 (channel 2) and 34 (channel 7) were driven with a high-voltage amplitude (7.5 V) sinusoid of 11.3 MHz for approximately 100 μs. a) Fourier transform plots of the optical modulation for MZI 34 and b) MZI 31, showing various harmonics and the location of applied digital frequency filters. c) Time-domain plots of the optical modulation for MZI 34 and d) MZI 31, showing the amplitudes of the filtered harmonics. MZI 34 shows stable modulation of the different frequency components at the expected harmonics, while MZI 31 shows many more frequency components, including a subharmonic at 5.65 MHz, whose amplitudes are drifting over time.

## 2. Supplementary Discussion

**Improvements in switching circuit design.** While the basic design of the photonic switch (Fig. 3b in the main text) is efficient in terms of number of optical channels relative to the number of optical modulators, the pulses from each output overlap significantly in time domain (Fig. 3d and Fig. 5a in the main text). If the photonic switch is used primarily for pulse picking applications, the effects of the overlapping channels may not be as important. However, if the photonic switch is to be used for pulse carving applications, then an improved design is warranted. Supplementary Fig. 5 shows a modified photonic switch, with an additional high-frequency MZI at the input. This MZI, labeled 00, provides an additional frequency component to the overall switching signals, narrowing all pulses at the output (Supplementary Fig. 5c). We see that narrower pulses are possible with a trade-off of more modulators and control complexity.

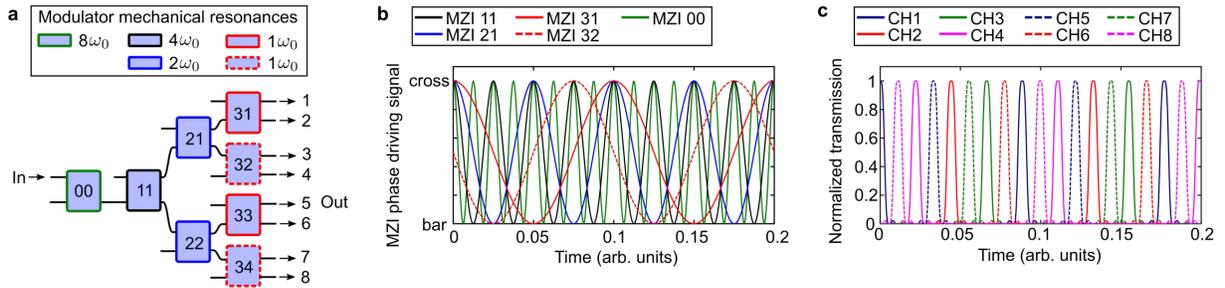

**Supplementary Fig. 5: Design improvements to resonantly actuated 1 x *N* photonic switches.** a) Schematic of a modified 1 x 8 photonic switch, showing a MZI 00 operating at 8 times the fundamental frequency added at the beginning of the switch's input. b) Plot of an example set of control signals for a subset of the MZIs, including the high-frequency MZI 00. c) Simulated optical outputs of the modified switch design, showing the effects of MZI 00 creating narrower pulses that have much less overlap in time compared to those without MZI 00.